\documentclass[]{trbunofficial}
\usepackage{graphicx}
\usepackage{amsmath,amssymb,amsfonts}
\usepackage{textcomp}
\usepackage{xcolor}
\usepackage[colorinlistoftodos,prependcaption,textsize=tiny]{todonotes}

\usepackage{algorithm}
\usepackage{algorithmic}

\usepackage{caption}
\usepackage{subcaption}
\usepackage[hidelinks]{hyperref}

\AuthorHeaders{Khoshkhah, Pourmoradnasseri, Ben Yahia, and Hadachi}
\title{A Random Walk Approach for Simulation-Based Continuous Dynamic Traffic Assignment}

\author{%
  \textbf{Kaveh Khoshkhah}\\
  kaveh.khoshkhah@ut.ee, ORDID: 0000-0003-4777-804X\\
\textit{ITS Lab, Institute of Computer Science}\\
\textit{University of Tartu, Tartu, Estonia} \\
  \hfill\break
  \textbf{Mozhgan Pourmoradnasseri, Corresponding Author}\\
    mozhgan@ut.ee, ORCID: 0000-0002-2092-816X\\
    \textit{ITS Lab, Institute of Computer Science}\\
    \textit{University of Tartu, Tartu, Estonia} \\
  \hfill\break%
    \textbf{Sadok Ben Yahia}\\
    sadok.ben@taltech.ee, ORCID: 0000-0001-8939-8948\\
    \textit{Data Science Group, Department of Software Science}\\
    \textit{Tallinn University of Technology, Tallinn, Estonia} \\
  \hfill\break%
  \textbf{Amnir Hadachi}\\
      hadachi@ut.ee, ORCID: 0000-0001-9257-3858\\
    \textit{ITS Lab, Institute of Computer Science}\\
    \textit{University of Tartu, Tartu, Estonia} 
}

\begin{document}
\maketitle

\section{Abstract}

This paper presents a new simulation-based approach to address the stochastic Dynamic Traffic Assignment (DTA) problem, focusing on large congested networks and dynamic settings. The proposed methodology incorporates a random walk model inspired by the theoretical concept of the \textit{equivalent impedance} method, specifically designed to overcome the limitations of traditional Multinomial Logit (MNL) models in handling overlapping routes and scaling issues. By iteratively contracting non-overlapping subnetworks into virtual links and computing equivalent virtual travel costs, the route choice decision-making process is shifted to intersections, enabling a more accurate representation of travelers' choices as traffic conditions evolve and allowing more accurate performance under fine-grained temporal segmentation.

The approach leverages Directed Acyclic Graphs (DAGs) structure to efficiently find all routes between two nodes, thus obviating the need for route enumeration, which is intractable in general networks. While with the calculation approach of downstream node choice probabilities, all available routes in the network can be selected with non-zero probability.  

To evaluate the effectiveness of the proposed method, experiments are conducted on two synthetic networks under congested demand scenarios using Simulation of Urban MObility (SUMO), an open-source microscopic traffic simulation software. The results demonstrate the method's robustness, faster convergence, and realistic trip distribution compared to traditional route assignment methods, making it an ideal proposal for real-time or resource-intensive applications such as microscopic demand calibration. 

\hfill\break%
\noindent\textit{Keywords}: Microscopic dynamic traffic assignment, Random walk, Equivalent impedance method, Traffic simulation, SUMO.
\newpage

\section{Introduction}

Traffic assignment is an extensively researched area in transportation research, yet various challenges remain unsolved in practice, particularly when dealing with large congested networks and dynamic settings \cite{peeta2001foundations,chiu2011dynamic}.

Dynamic traffic assignment (DTA) is essential for generating time-dependent traffic models by allocating demand to road networks. This step replicates the decision-making process of individual travelers, leading to a more accurate depiction of vehicle interactions within the system \cite{zhang2011behavioral}. DTA is particularly important to model and predict the movement of vehicles through a transportation network as traffic conditions change over time. It is immensely useful for understanding the evolution of congestion patterns and evaluating the impact of various transportation management strategies.

The main approaches for the DTA problem in the literature include analytical methods \cite{lu2022analytical, lu2018probabilistic} and simulation-based methods \cite{shafiei2018calibration, duell2016deployment} for the network loading process. 
Both approaches are founded on Wardrop's user equilibrium (UE) and system optimal (SO) principles. Under the principle of UE, no vehicle can reduce its travel cost by unilaterally changing to another route. This principle assumes that each vehicle aims to minimize its own travel cost by following a selfish routing strategy. In contrast, the principle of SO considers minimizing the average (or total) travel cost in networks. Under this principle, each vehicle selects a route that minimizes its travel cost and contributes to reducing the overall travel time of the entire network.

Simulation-based methods for traffic assignment aim to approximate UE or SO iteratively \cite{sbayti2007efficient, mehrabani2022proposing, mehrabani2023multiclass}. In practice, the number of iterations for converging to equilibrium conditions can be very high for congested networks. Thus, the traffic assignment method's time complexity and convergence speed directly impact the overall running time of traffic simulations. In calibration methods for demand estimation using real-world measurements, executing multiple traffic assignments is necessary \cite{kaveh2022, pour2022}. Hence, minimizing the computational burden of the traffic assignment step and ensuring the method's robustness significantly impact the results, especially in real-time applications \cite{khoshkhah2022}.

For a given demand, the DTA problem consists of two main steps: finding the alternative route set between origin-destination (OD) pairs and implementing the route choice model for network loading.

It is important to note that enumerating all routes between a pair of nodes in a general network is known to be an intractable (NP-hard) problem. Consequently, the process of selecting the alternative route set involves computational challenges, and finding an exhaustive set of routes may not be well-defined in large networks. To address this, various methods for the selection of ``reasonable'' alternative routes have been proposed to obviate route enumeration \cite{dial1971probabilistic, li2005algorithm}. Some practical methods involve static selection, where the choice set is predetermined. For instance, external data sources like the Google Maps Directions API can provide a fixed number of plausible routes per OD pair and at different times of the day \cite{cabannes2023creating,50831}. Alternatively, a fixed number of time-dependent shortest paths between each OD pair can be chosen based on iterative DTA simulations \cite{pour2022}.

Based on existing literature and research in the field, route choice models commonly rely on a predefined set of alternative routes. This route set is typically denoted as a subset $\mathcal{C}$ of the universal set of all routes $\mathcal{U}$ available in the transportation network. Regardless of the approach for selecting the alternative route set, it is essential to recognize that the probability of choosing a route $r$ is conditioned on the route set \cite{frejinger2009sampling}:

\begin{equation}
P(r) = \sum_{\mathcal{C}\subseteq \mathcal{U}} P(r\vert \mathcal{C}) P(\mathcal{C}).
\end{equation}

The multinomial logit model (MNL) for route choice is a common method used to distribute trips stochastically among a given set of alternative routes $\mathcal{C}$ between the OD pair $(o,d)$ \cite{han2003dynamic, osorio2019dynamic,arora2021efficient, mehrabani2022proposing}. This route choice model incorporates a dispersion parameter $\gamma$ and route cost $\theta_a$ for each route $a$.

\begin{equation} \label{eq:logit}
P(r\vert \mathcal{C})=\frac{\exp (\gamma \theta_r)}{\sum_{a\in \mathcal{C}} \exp (\gamma \theta_a)}.
\end{equation}

Logit models, while possessing known limitations in addressing overlapping routes, have gained widespread practical appeal due to their straightforward closed-form expression for choice probabilities \cite{gao2008adaptive, SUMO2018}.

This study presents a novel random walk approach for addressing the simulation-based stochastic DTA problem that aims to converge to user equilibrium (UE). The proposed route choice model draws inspiration from the theoretical concept of the \textit{equivalent impedance} method, previously introduced to tackle the overlapping and scaling deficiencies of the MNL model \cite{li2015modeling, li2019modelling}. The method involves iteratively contracting non-overlapping subnetworks into virtual links and computing equivalent virtual travel costs for these subnetworks, effectively shifting the route choice decision-making process to intersections from the origins. In other words, at each time interval $\Delta$, when a driver is at node $i$ and intends to travel to destination $d$, the probability of selecting a downstream node $j$, denoted as $P_{ij}^{\Delta d}$, is calculated. Importantly, node $i$ can represent either the origin node or any internal node in the trip trajectory. This approach enables the execution of a truly dynamic and continuous traffic assignment, wherein the downstream choice probabilities are updated accordingly if the driver remains in the middle of the trip and advances to the next time interval $\Delta+1$. In other words, the driver's knowledge of the path is updated in our method if the trip is unfinished in a time interval. 

To overcome the challenge of route enumeration, the study initially computes the probabilities $P_{ij}^{\Delta d}$ for an acyclic subnetwork derived from the original transportation network. Subsequently, the removed edges with sufficiently small choice probabilities are reintroduced into the model. Although the route enumeration problem is known to be intractable in general networks, Directed Acyclic Graphs (DAGs) present a different scenario. DAGs, which are directed networks devoid of directed cycles, enable the efficient identification of all paths between two nodes. The absence of cycles ensures the absence of infinite loops or redundant paths to consider. Moreover, a general directed network can be transformed into a DAG by selectively removing certain links to break cycles, a process known as the feedback arc set problem (FAS). While the FAS problem is NP-complete in the general case, it can be solved polynomially for planar networks, such as transportation networks \cite{bang2008digraphs, baharev2021exact}.

We would like to clarify that \cite{frejinger2009sampling} introduced  a random walk model as a method for route generation to estimate choice set bias and route sampling correction within the MNL model. In their approach, the model initiates from a fixed origin, and at each step, the next link is sampled based on its distance from the shortest path to the destination. It is essential to note that their use of static random walk probabilities is specifically intended for route generation and not route assignment. 

In the context of DTA, the proposed random walk model assumes that individual vehicles make routing decisions at intersections. At each step of the algorithm, the driver probabilistically selects one of the downstream nodes based on the latest estimated travel costs in the induced subnetworks. As a result, the probability of choosing a route $r$ between origin $o$ and destination $d$ can be calculated as $\Pi_{(i,j)\in r} P_{ij}^{\Delta^t d}$, where $\Delta^t$ represents the corresponding time frame. This dynamic behavior enables a more realistic representation of travelers' route choices in response to changing traffic conditions and evolving travel costs at different time intervals.

Furthermore, in conjunction with the route enumeration approach based on a DAG, the proposed model can generate all possible routes in the network with a non-zero probability, thereby addressing the limitations of existing dynamic models that often focus on only a small fraction of available routes.

The efficacy of the proposed methodology is evaluated on two synthetic networks under congested conditions. The source codes are made publicly available \footnote{\href{https://github.com/Khoshkhah/DTA}{https://github.com/Khoshkhah/DTA}} and integrated into Simulation of Urban MObility (SUMO) \cite{SUMO2018}, a commonly used open-source simulation tool. The results of experiments demonstrate that our method exhibits a more robust performance, faster convergence, and a more realistic trip distribution. Furthermore, our route assignment method can be executed in parallel, reducing the running time in calibration and real-time applications.

In the subsequent sections, we first describe the equivalent impedance method for calculating virtual travel costs. We then demonstrate how downstream choice probabilities are computed, initially starting from simple network structures and gradually expanding the concept to a general network. Subsequently, we present the results obtained from our implementations on two synthetic networks and conduct a comparative analysis with the performance of SUMO's built-in methods. Finally, we conclude by outlining potential directions for future research.

\section{Equivalent Impedance Method for Virtual travel cost Calculation}
In this section, we explain a variation of the logit model for tackling the issue of overlapping routes and scaling in networks. Then,  we use the concepts of virtual links and virtual travel cost in the proposed method. 

Various modifications of the logit model are proposed in the literature to improve the route choice model more realistically. Inspired by the concept of equivalent impedance, an improved version of the logit model is proposed in \cite{li2015modeling, li2019modelling} to address the overlapping and scaling issues in the traditional approach, to only use the non-overlapping portions of the routes in the travelers' decision making (check \cite{li2015modeling} for examples). If the non-overlapping part of paths between $o$ and $d$ falls only in the subnetwork between node $i$ and node $j$, with the minimum travel cost $\pi_{ij}$, the probability of choosing route $r_{ij}$ connecting $i$ and $j$ is equivalent to 

\begin{equation} \label{eq:modif_logit}
    P_{r_{ij}}=\frac{\exp ( \gamma \theta_{r_{ij}}/\pi_{ij})}{\sum_{a\in R_{ij}} \exp ( \gamma \theta_a/\pi_{ij})}.
\end{equation}

The primary enhancement of Equation \ref{eq:modif_logit} compared to the classical MNL model is its exclusion of the overlapping portions of routes in decision probability when selecting a route between origins $o$ and destinations $d$. This probability calculation can be used differently for calculating the equivalent travel cost of the subnetwork consisting of the nonoverlapping routes connecting nodes $i$ and $j$ by introducing \textbf{virtual link} $\overset{\sim}{(i,j)}$ as an equivalence of the subnetwork. The virtual link is strategically constructed to ensure that the probability of drivers selecting the original subnetwork equals that of opting for the virtual link $\overset{\sim}{(i,j)}$, assuming a route choice framework. 

Additionally, the corresponding \textbf{virtual travel cost}  $\theta_{\overset{\sim}{(i,j)}}$ is defined as the virtual travel cost of the original subnetwork.

Therefore, if $n$ parallel paths connect $i$ and $j$, by replacing  $P_{\overset{\sim}{(i,j)}} = 1$, in equation \ref{eq:modif_logit},  the virtual travel cost $\theta_{\overset{\sim}{(i,j)}}$ will be calculated according to 

\begin{equation} \label{eq:equiv-time}
\theta_{\overset{\sim}{(i,j)}} = \frac{\pi_{ij}}{\gamma} \ln \left[ \sum_{k=1}^n \exp \left( \gamma \frac{\theta_k}{\pi_{ij}} \right) \right].
\end{equation}

The calculated virtual travel costs exhibit similar properties to normal travel costs. For a single link, the virtual and normal travel costs are identical. The virtual travel cost of a route can be obtained by summing the virtual or normal travel costs of its constituent links.

\begin{figure}[!tbh]
     \centering
     \begin{subfigure}[b]{0.7\textwidth}
         \centering
         \includegraphics[width=\textwidth]{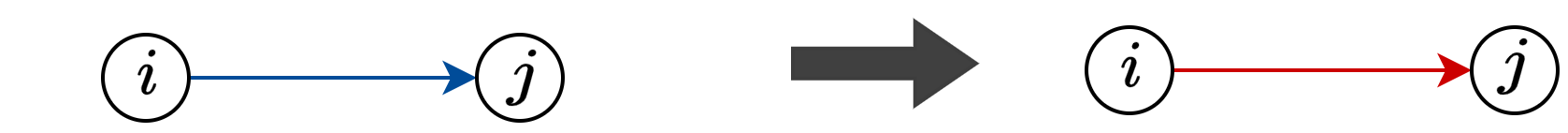}
         \caption{}
         \label{fig:a-link}
     \end{subfigure}
     \hfill \\
     \begin{subfigure}[b]{0.7\textwidth}
         \centering
         \includegraphics[width=\textwidth]{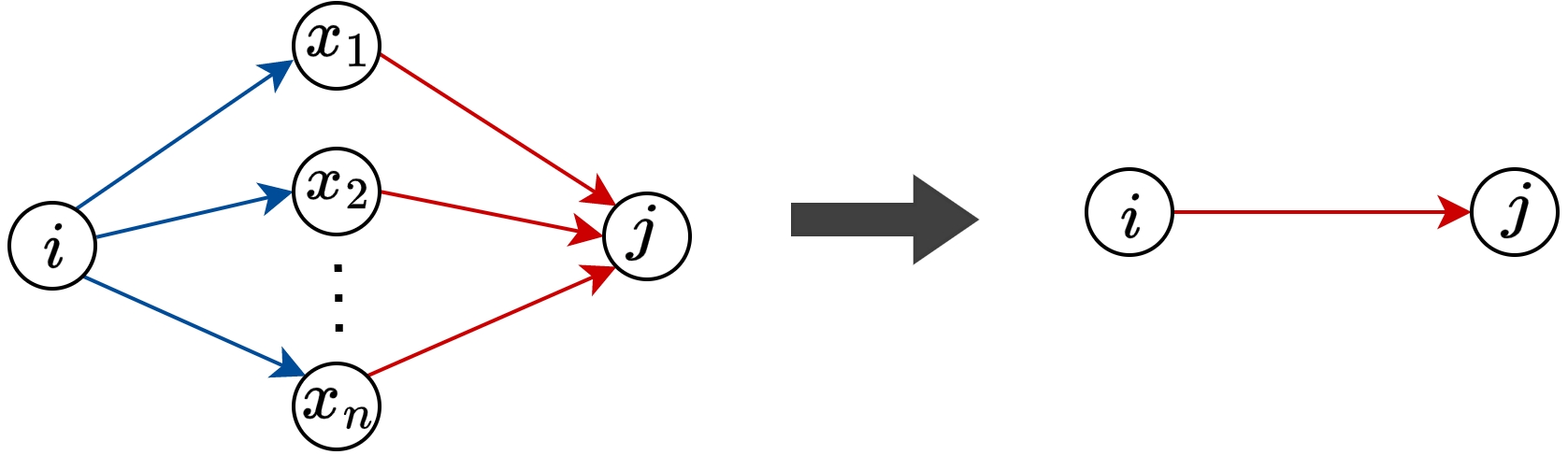}
         \caption{}
         \label{fig:b-link}
     \end{subfigure}
    \caption{Base cases for forming virtual links from real links, for route choice from $i$ to $j$. Blue represents real links, and red represents virtual links.}
    \label{fig:EQ-Imp}
\end{figure}

Figure \ref{fig:EQ-Imp} illustrates the iterative transformation of non-overlapping subnetworks into equivalent links using this methodology.

In Figure \ref{fig:a-link}, the probability of moving from node $i$ to node $j$ is trivially equal to one, and the travel cost from $i$ to $j$ in the virtual link is the same as that in the real link. The virtual travel cost of the subnetwork shown in Figure \ref{fig:b-link} is calculated based on Equation \ref{eq:equiv-time}.

The process continues iteratively, replacing subnetworks without overlapping links with virtual links until there are only parallel non-overlapping routes remaining between a given pair of nodes. Then, the probability of choosing a subroute $r$ between nodes $i$ and $j$ is given by 

\begin{equation} \label{eq:downs}
    P({r\vert (i,j)})=\frac{\exp ( \gamma \theta_{r_{s}}/\pi_{ij})}{\sum_{a\in R_{ij}} \exp ( \gamma \theta_a/\pi_{ij})}.
\end{equation}

Equation \ref{eq:downs} is equivalent to Equation \ref{eq:modif_logit}, with the only difference being that the travel costs $\theta$ can be virtual. Finally, the probability of choosing route $r$ in the network will be

\begin{equation} \label{eq:downs_1}
    P(r) = P({r\vert (i,j)})\cdot P(\overset{\sim}{(i,j)}),
\end{equation}
when $P(\overset{\sim}{(i,j)})$ represents the probability of choosing the virtual link  $\overset{\sim}{(i,j)}$.

\section{Calculating Downstream Node Choice Probabilities}

This section explains the steps involved in the proposed random walk route choice model for calculating downstream node choice probabilities. The main objective is to compute the probability $P^{\Delta d}_{ij}$, which represents the likelihood of selecting node $j$ during the time interval $\Delta$, given that the driver is at node $i$ and intends to travel to destination $d$, while there exists a link connecting nodes $i$ and $j$.
To begin, we introduce some structures from graph theory. Next, we compute the choice probability for a specific directed network structure. Finally, we present the algorithm for the general case. 

An \textbf{out-tree} is a directed rooted tree structure where each node, except the root node, has an in-degree of one. The root node has an in-degree of zero. A \textbf{modified out-tree} is an out-tree with an ``extra-node'' that can have an in-degree greater than one, allowing for multiple incoming edges.

An \textbf{in-tree} is similar to an out-tree, but the directions of edges are reversed. A \textbf{modified in-tree} is an in-tree with an ``extra-node'' that can have an out-degree greater than one.

\begin{figure}[htbp]
     \centering
     \begin{subfigure}[b]{0.2\textwidth}
         \centering
         \includegraphics[width=\textwidth]{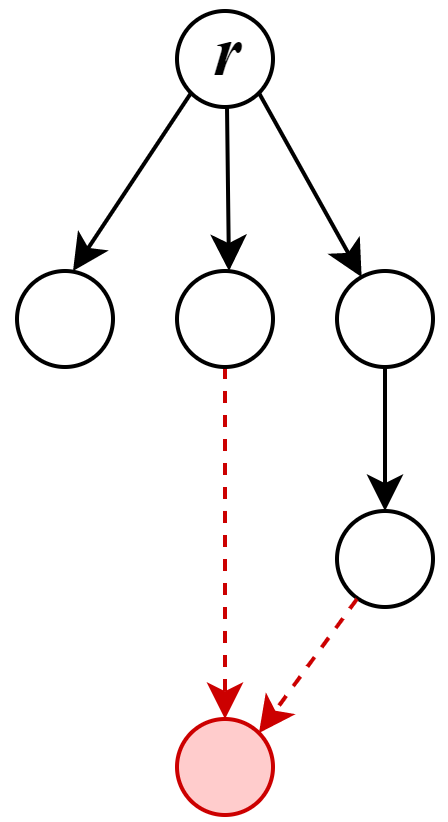}
         \caption{}
         \label{fig:out}
     \end{subfigure}
        \hfill
     \begin{subfigure}[b]{0.2\textwidth}
         \centering
         \includegraphics[width=\textwidth]{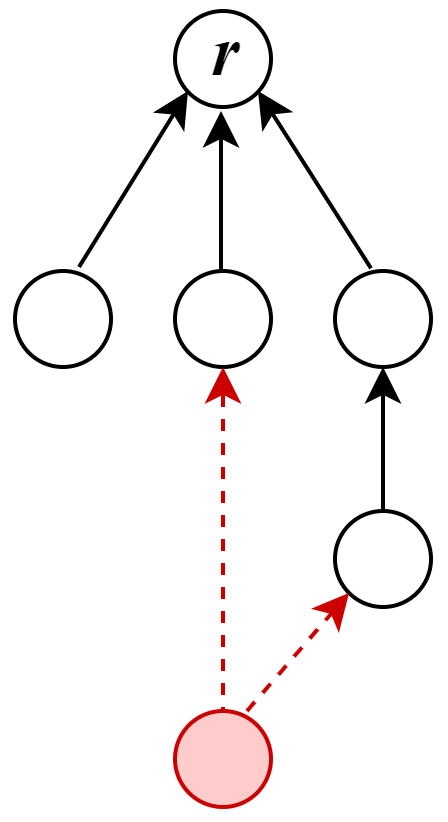}
         \caption{}
         \label{fig:in}
     \end{subfigure}
       \hfill
    \begin{subfigure}[b]{0.24\textwidth}
         \centering
         \includegraphics[width=\textwidth]{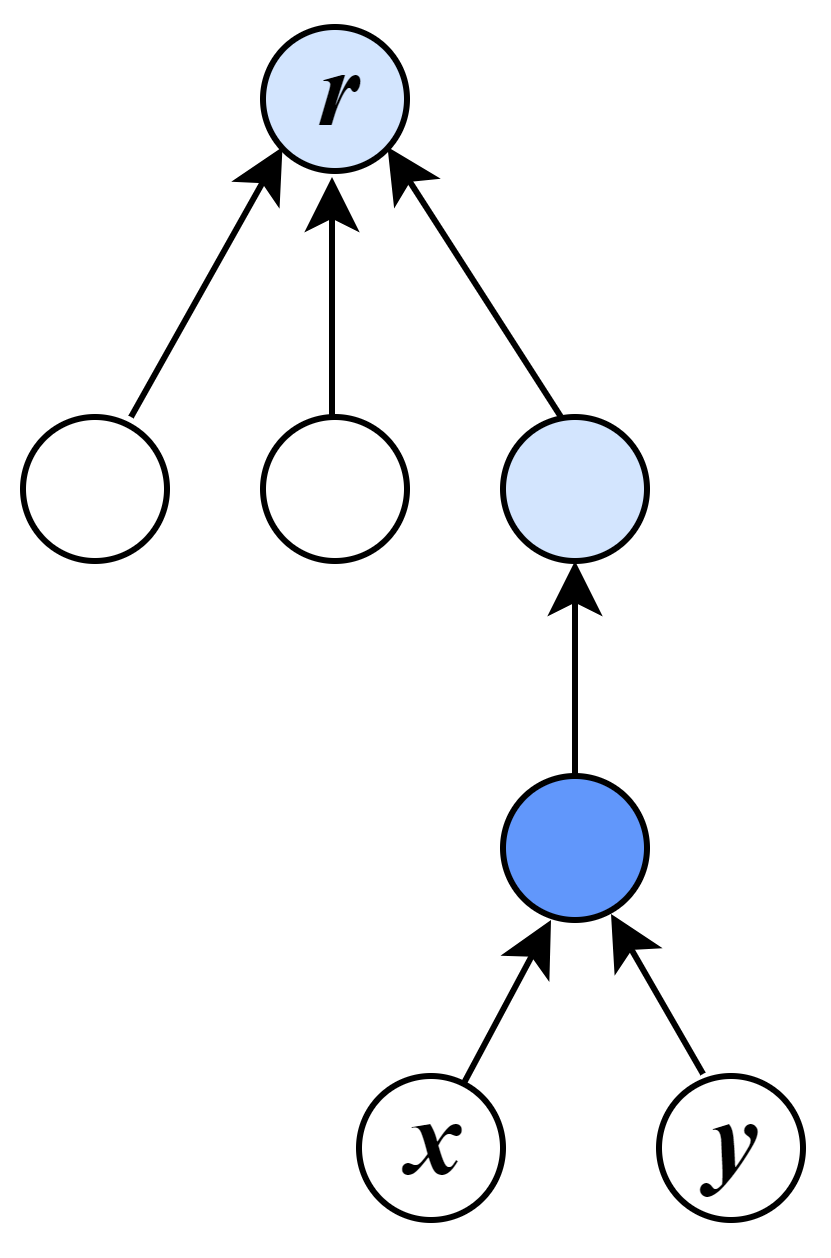}
         \caption{}
         \label{fig:LCA}
     \end{subfigure}
    \caption{(a) shows an out-tree, and (b) displays an in-tree in black. The modified out-tree (in-tree) is formed by adding an extra node in red. In (c), the common ancestors of nodes $x$ and $y$ are in blue, and their LCA is highlighted in dark blue.}
    \label{fig:in-out}
\end{figure}

\textbf{Lowest Common Ancestor (LCA)} for a subset of nodes in an in-tree is the node located farthest from the root and is a common ancestor to all the nodes in the given subset. See Figure \ref{fig:in-out} for illustration and check \cite{mehlhorn2008algorithms} for more detailed information.

\subsection{Calculating downstream node choice probability in a modified out-tree}

As a special case, we first discuss computing $P^{\Delta d}_{ij}$ for a modified out-tree, where $d$ serves as the extra node and $r$ is the root of the tree.

The primary advantage of the modified out-tree, in our context, is that for every node $i$ in the tree when $i\neq r \text{ or } d$, there exists exactly one path from $r$ to $d$, containing $i$. The probability of selecting this path is calculated by multiplying the probabilities of selecting the downstream nodes.

Algorithm \ref{alg:out-tree} outlines the steps for calculating downstream node choice probabilities and virtual travel costs to $d$ for all internal nodes of an out-tree when $d$ is the destination.

\vspace{.5cm}
 \begin{algorithm}
 \caption{Downstream Node Choice Probability and Virtual Link travel cost\\ in Modified Out-Tree}
 \begin{algorithmic}[1] \label{alg:out-tree}
 \renewcommand{\algorithmicrequire}{\textbf{Input:}}
 \renewcommand{\algorithmicensure}{\textbf{Output:}}
 \REQUIRE Modified out-tree with root $r$ and extra node $d$,  Link travel costs
 \ENSURE  Downstream node choice probabilities $P^{\Delta d}_{ij}$,  Virtual travel cost $\theta_{\overset{\sim}{(r,d)}}$

\STATE Perform a topological sort $\mathcal{S}$ from $r$ to $d$
\FOR {each node $i$ in the reversed order of $\mathcal{S}$}
    \STATE Calculate downstream node choice probabilities at $i$ (Equation \ref{eq:downs})
    \STATE Calculate the virtual link travel cost from $i$ to $s$ (Equation \ref{eq:equiv-time})
    \STATE Replace the subnetwork with the virtual link (Figure \ref{fig:EQ-Imp})
\ENDFOR
\end{algorithmic} 
\end{algorithm}

\subsection{Calculating downstream node choice probability in a modified in-tree}

A similar approach to Algorithm \ref{alg:out-tree} cannot be directly applied to a modified in-tree with the root $r$ as a destination. This is because, during the process of reducing subnetworks to virtual links, there is a possibility of overlapping subnetworks. In the case of a modified in-tree, our goal is to calculate the probabilities of selecting a downstream node within the extra node $o$. To address this, we reverse the direction of the edges in the modified in-tree, transforming it into a modified out-tree.

By obtaining the modified out-tree, we can determine the downstream node choice probabilities at $s$ by considering the probability of selecting the path from the root $r$ to $o$, which includes the specific downstream node in the modified out-tree. In the context of the in-tree, the virtual travel cost from $o$ to $r$ is equivalent to the virtual travel cost from the root $r$ to the extra node $o$ in the modified out-tree. This reversal of edge directions and corresponding virtual travel costs enables us to calculate the desired probabilities for the modified in-tree structure.



\vspace{.5cm}
\subsection{Calculating downstream node choice probabilities in a general network}

In this section, we present Algorithm \ref{alg:downstream_prob} for calculating downstream node choice probabilities in a directed network $G$ when considering the destination node $d$, by utilizing the concepts introduced in previous sections.

The algorithm operates on a directed acyclic subnetwork of $G$ (DAG) and traverses the nodes based on a specified topological order $\mathcal{S}$ \cite{pang2015topological}. The topological order can be obtained by performing the Dijkstra algorithm from the destination node $d$ to all other nodes when the directions of links are reversed.  At each step, it calculates the virtual travel cost and downstream node choice probabilities from a vertex $i$ to $d$, while maintaining a subnetwork $T_{\text{virtual}}$ with an in-tree structure.
For each node $i$ in $\mathcal{S}$, the algorithm identifies the LCA $l$ of its downstream nodes within $T_{\text{virtual}}$. A sub in-tree $T_{\text{sub}}$ is created from $T_{\text{virtual}}$ using $l$ as the root. The modified in-tree $G_{\text{modified}}$ is then obtained by adding node $i$ to $G_{\text{sub}}$.

Next, the algorithm calculates the probability of choosing each downstream node of $i$ in $T_{\text{modified}}$ for the destination node $l$. Additionally, it computes the estimated travel cost for the virtual link from $i$ to $l$. Finally, $T_{\text{virtual}}$ is updated by adding the virtual link from $i$ to $l$ along with its corresponding travel cost value.

\begin{algorithm}
\caption{Downstream Node Choice Probabilities}
\label{alg:downstream_prob}
\begin{algorithmic}[1]
\renewcommand{\algorithmicrequire}{\textbf{Input:}}
\renewcommand{\algorithmicensure}{\textbf{Output:}}
\REQUIRE Directed network $G$, destination node $d$
\ENSURE Downstream node choice probabilities $P^{\Delta d}_{ij}$ at each node $i$ for destination $d$
\STATE Acquire a topological sort $\mathcal{S}$ of nodes based on the shortest path to $d$ in ascending order; 
\STATE Create a DAG $A$ based on the topological sort $\mathcal{S}$, from $G$; 
\STATE Initialize an in-tree $T_{\text{virtual}}$ with node $d$ as root\;
\FOR{each node $i$ in $\mathcal{S}$}
    \STATE $D_i:=$ set of downstream nodes of $i$ in $A$
    \STATE Find LCA $l$ of $D_i$ in $T_{\text{virtual}}$
    \STATE Create a sub in-tree $T_{\text{sub}}$ from $T_{\text{virtual}}$ with root $l$
    \STATE Form a modified in-tree $T_{\text{modified}}$ by adding node $i$ to $T_{\text{sub}}$ and connecting $i$ to nodes in $D_i$
    \STATE Calculate the probability of choosing each downstream node of $i$ in $T_{\text{modified}}$\\ for destination $l$
    \STATE Calculate the virtual travel cost for the virtual link from $i$ to $l$ 
    \STATE Update $T_{\text{virtual}}$ by adding the virtual link from $i$ to $l$ with its virtual travel cost value
\ENDFOR
\end{algorithmic}
\end{algorithm}

More precisely, Algorithm \ref{alg:downstream_prob} computes the downstream node choice probabilities in a DAG, which excludes some links of the original graph. To extend the choice to the whole network links, we assign a small probability of $\beta = 0.1$ for selecting these excluded links and then normalize the probability distribution at each node. The parameter $\beta$ should be selected in such a way that it is less than the minimum downstream node choice probability, while still not being too small to approach zero.



\section{Microscopic Random walk algorithm for DTA}

The presented methodology for random walk DTA is tested at the microscopic level with the use of SUMO \cite{SUMO2018} tool, an open-source microscopic traffic simulation software. SUMO is suitable for conducting microscopic traffic simulations in academic and open-source settings. As open-source software, it offers cost-effective and efficient ways to understand traffic behavior at a high level of granularity. 

\begin{figure*}[h]
\centerline{\includegraphics[width=1.1\textwidth]{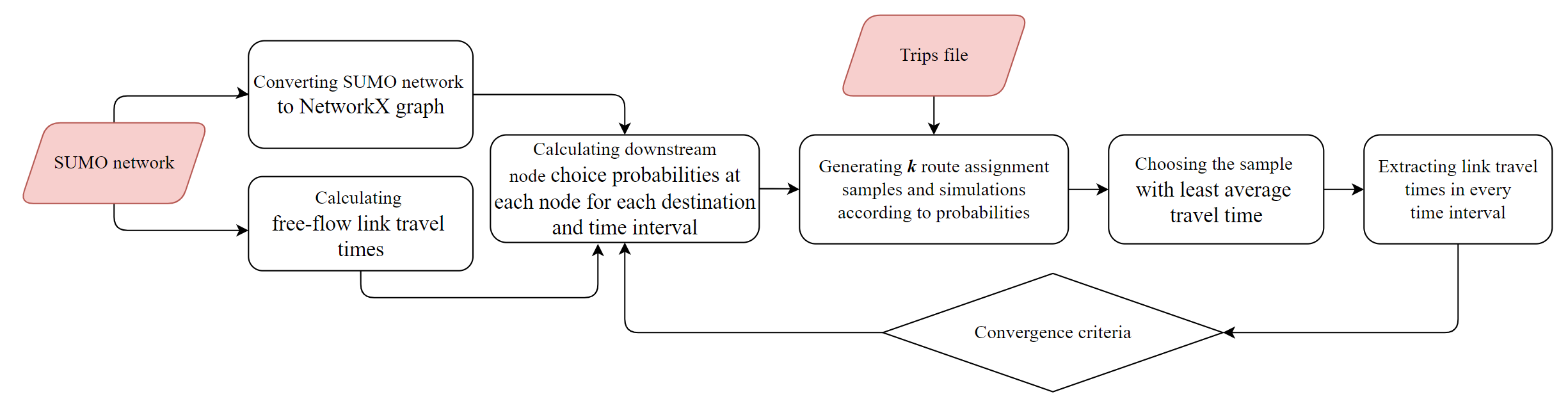}}
\caption{Flowchart of the general methodology for the random walk model.}
\label{fig:diagram}
\end{figure*}

The general flowchart of the algorithm is illustrated in Figure \ref{fig:diagram}. The algorithm takes two inputs: the network and the demand. These inputs are provided as a SUMO network and a trip file. Each trip is defined by the origin node, destination node, and departure time. 

After computing the probabilities of downstream node choice at each node, for every destination, and in specific time intervals, a route is assigned to each trip using a random walk on the network nodes, following the probability distribution.  

In each iteration, the algorithm performs $k$ route assignments and simulations in parallel and then selects the best route assignment based on the average link travel times. The algorithm terminates when the deviation of average travel times between successive iterations is less than a defined threshold, indicating that the user equilibrium condition is ``almost'' reached. Otherwise, the link travel times are updated, and the probabilities are recalculated for the next iteration.

The time interval for the probability distribution is determined by considering the trip's departure time and the estimated travel cost.
If the selected route contains loops, the algorithm proceeds to remove these loops to ensure the resulting routes are loop-free.
Additionally, to ensure the convergence of the algorithm in the iterative process, we apply the computed probabilities as weights to the previous probabilities in each iteration and normalize the results. This approach helps maintain the stability of the method as the system reaches a user equilibrium condition and travel costs converge. Then, the calculated probabilities remain unchanged, contributing to the robustness of the method.

\section{Experiments and Results} 

The proposed methodology has been tested on two distinct synthetic networks: a small grid and a medium-sized network with a random structure. The evaluation involves a comprehensive analysis of the total travel time and total travel distance of simulated routes across multiple iterations for both networks under congested demand scenarios. These results are then compared with the logit and Gawron route choice models implemented in the \textit{duaIterate.py} methods.

All experiments were conducted on a machine equipped with an Intel Core i7-9800X sixteen-core processor and 64 GB of RAM, running on Ubuntu. For both experiments, $k = 16$ (equal to the number of cores), enabling route assignment sampling and simulations to be generated in parallel in each iteration, thereby optimizing computational efficiency.

In the following sections, we briefly explain the route assignment steps implemented in the SUMO tool in order to have a better comparison with the proposed method. Subsequently, a method for converting SUMO network to a normal network is presented. Finally, we present the detailed results obtained from applying the proposed method to the two sample networks. 


\subsection{Route-Choice Algorithm in SUMO}

In SUMO, the computation of DTA is performed using the \textit{duaIterate.py} procedure, which follows an iterative approach by alternating between calculating travel times from simulations and performing route assignments based on these calculated travel times \cite{sumo-dynamic-user-assignment}. The following steps outline the main process involved in generating a microscopic simulation scenario based on the given Origin-Destination (OD) demand: 
\begin{enumerate}
    \item \textit{Network generation}: The first step involves creating the network structure for the simulation.
    \item \textit{Conversion of OD demand to trips}: The OD demand data is converted to individual trips using the OD2trip function. Each trip consists of the origin and destination links and the trip start time.
    \item \textit{DUA computation}: The \textit{duaIterate.py} procedure is executed to compute the DUA. It iterates for a defined number of steps, performing the following sub-steps:
    \begin{enumerate}
        \item \textit{duarouter shortest route generation}: The duarouter module is used to generate the shortest routes between OD pairs for each vehicle. This is done based on edge weights, which represent travel times. The process starts with an empty-network assignment.
        \item \textit{Alternative route addition}: The shortest route obtained for each OD pair is added to the set of alternative routes. A predefined maximum number of routes to maintain is specified, with the default being five.
        \item \textit{Route choice selection}: Each driver selects a route from the list of alternative routes, following a probability distribution computed in the route choice model. This model can be based on either Gawron or logit algorithms.
        \item \textit{Simulation and travel time update}: A simulation is run for the selected routes to determine the current travel times for each route. The mean travel times for edges are then updated based on the simulation results.
        \item \textit{Convergence check}: The current mean route costs are compared with the previous values. If the mean travel time decreases below a certain threshold, it is considered that the algorithm has converged to user equilibrium, and the process exits.
    \end{enumerate}
    \item If the mean travel time is still above the threshold, new routes are computed using each vehicle's last known route costs. These new routes are added to the corresponding vehicle's set. The routes of all vehicles are then collected, and a route distribution mechanism (either Gawron or logit algorithms) is used to select the route for each vehicle in the next iteration.
\end{enumerate}

By following these steps, the \textit{duaIterate.py} procedure facilitates the computation of DTA by iteratively adjusting routes based on travel times until convergence to user equilibrium is achieved.



\subsection{Converting SUMO network to a normal network} \label{app:network}

\begin{figure}[h!]
     \centering
     \begin{subfigure}[b]{0.7\textwidth}
         \centering
         \includegraphics[width=\textwidth]{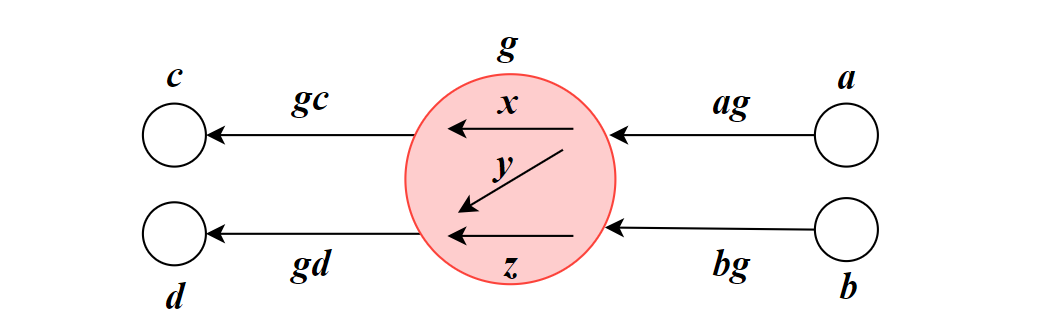}
         \caption{Sample SUMO node in red with internal links.\\}
         \label{fig:sumo-net}
     \end{subfigure}
    \\
     \begin{subfigure}[b]{0.7\textwidth}
         \centering
         \includegraphics[width=\textwidth]{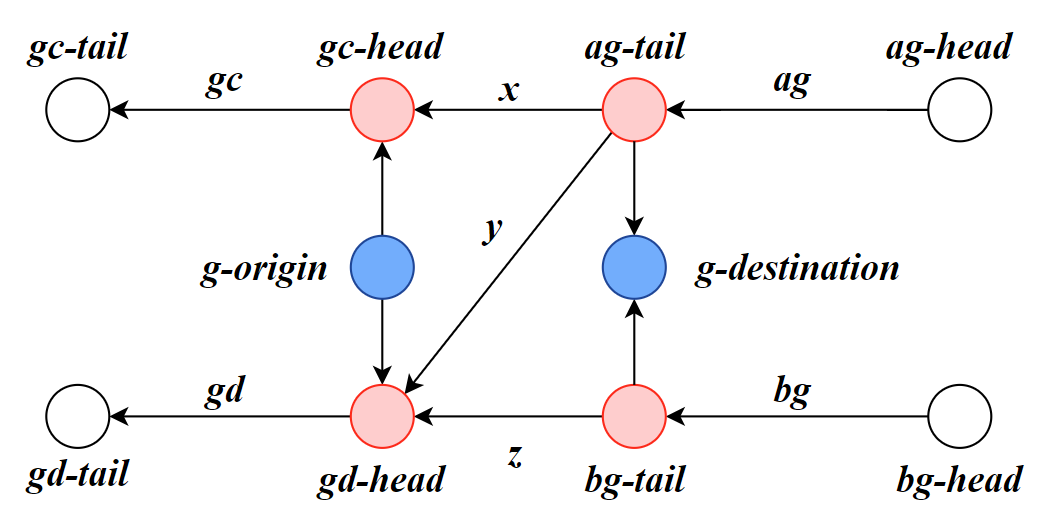}
         \caption{Converting internal links to normal directed network links.}
         \label{fig:network}
     \end{subfigure}
    \caption{Transforming SUMO network to a directed graph.}
    \label{fig:link}
\end{figure}

The SUMO network \cite{sumo-road} consists of ``normal links'' representing road segments and connections between two nodes, as well as ``internal links'' within intersections. These internal links connect an incoming normal edge with an outgoing normal edge and define the right-of-way. In Figure \ref{fig:sumo-net}, internal links are denoted as $x$, $y$, and $z$, while the remaining links are considered normal links.

For each node in the SUMO network, we replace the node and its internal edges with a subgraph. This replacement ensures a one-to-one correspondence between routes in SUMO and the directed network representation. Figure \ref{fig:network} illustrates the transformed version of the network shown in Figure \ref{fig:sumo-net}, where all links are normal. To maintain the one-to-one correspondence, additional links and vertices are introduced.

By performing this conversion, we can utilize graph theory algorithms and reach packages such as NetworkX \cite{networkx} in the subsequent steps. Routes are defined as a sequence of consecutive vertices within the transformed network.

\newpage

\subsection{Small Network Experiment}
\begin{figure}[h!]
\centerline{\includegraphics[width=0.2\textwidth]{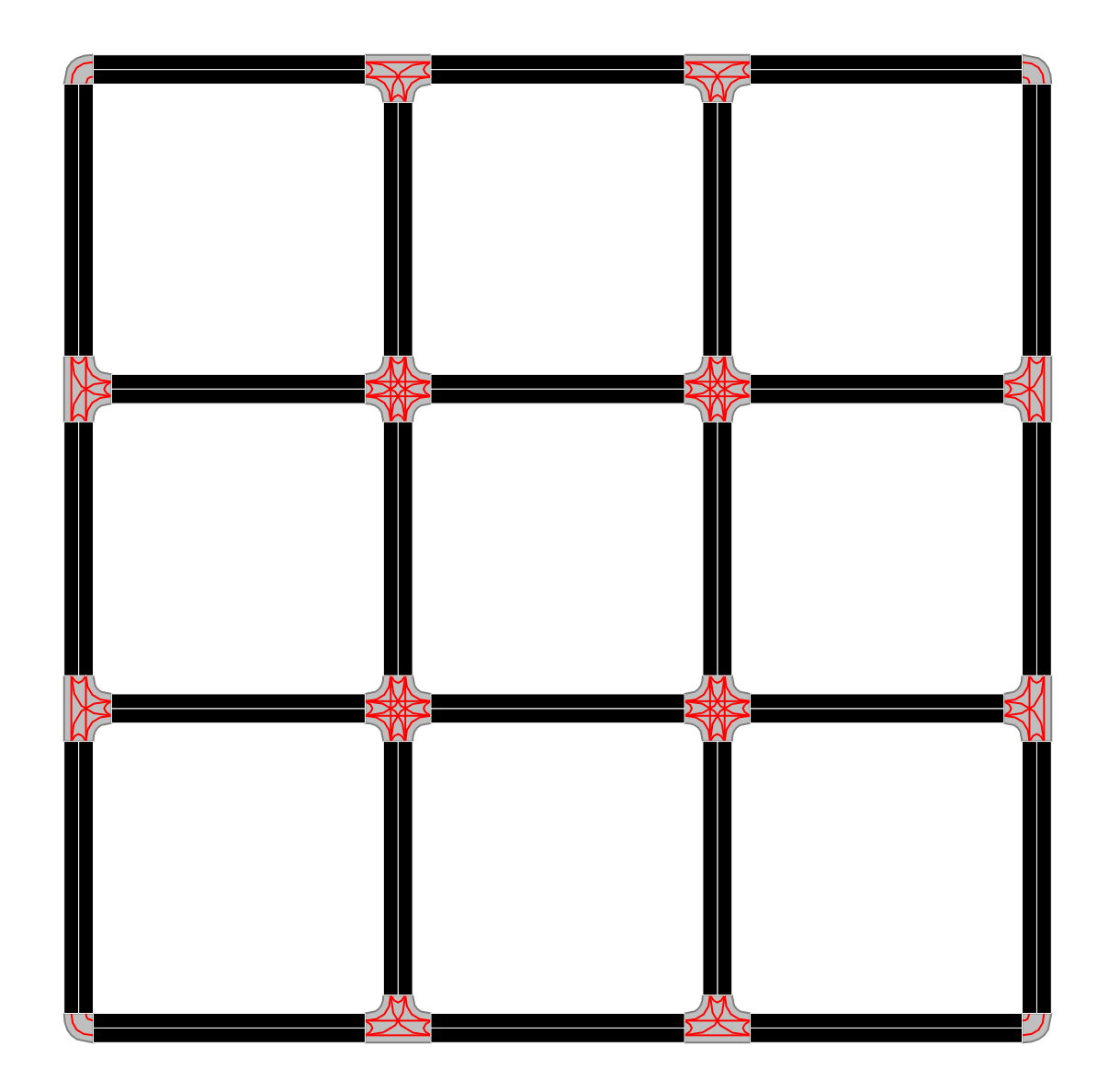}}
\caption{The structure of the small network.}
\label{fig:small-grid}
\end{figure}

\begin{figure}[H]
     \centering
     \begin{subfigure}[b]{0.47\textwidth}
         \centering
         \includegraphics[width=\textwidth]{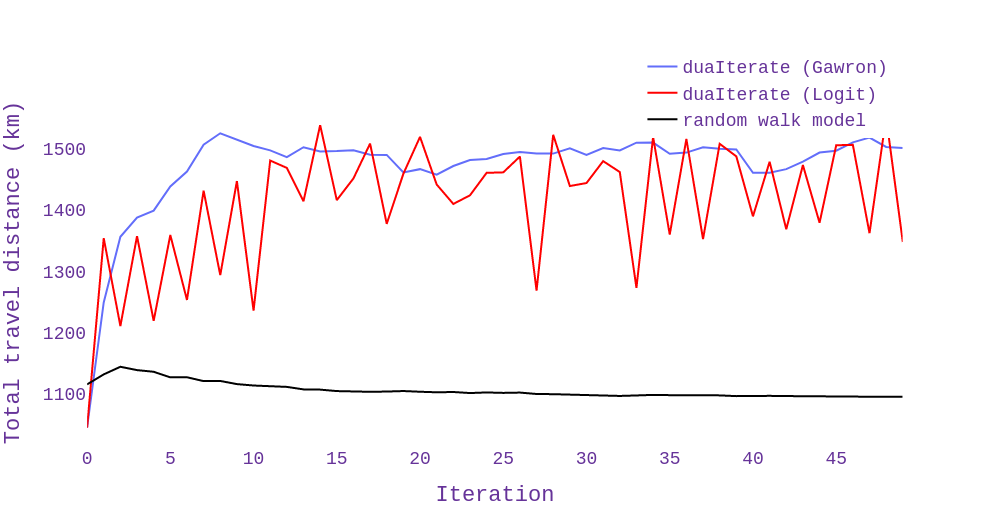}
         \caption{Total travel distance.}
         \label{fig:small-TD}
     \end{subfigure}
     \hfill
     \begin{subfigure}[b]{0.47\textwidth}
         \centering
         \includegraphics[width=\textwidth]{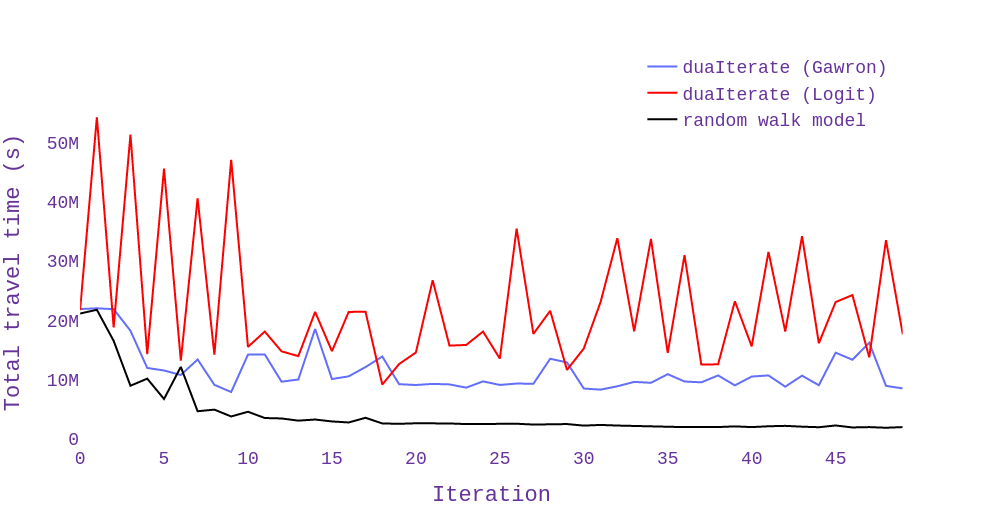}
         \caption{Total travel time.}
         \label{fig:Small-TT}
     \end{subfigure}
    \caption{Comparison of total travel time and distance for traffic assignment of the proposed method and \textit{duaIterate.py} in the grid network.} \label{fig:small-results}
\end{figure}

The structure of the small network is depicted in Figure \ref{fig:small-grid}. This network comprises 48 links and 16 junctions. Each link has a length of 400 meters and consists of a single lane. In this analysis, all nodes within the network are regarded as both origins and destinations. A random traffic demand of $8,013$ vehicles was generated for a one-hour simulation period, divided into four 15-minute time intervals to simulate traffic flow.
Figure \ref{fig:small-results} presents the performance metrics of the proposed route choice model compared to the logit and Gawron models in \textit{duaIterate.py}. The results of the random walk model show a considerable improvement in the total travel distance and time of simulated trips. Moreover, the method reaches the equilibrium condition much faster and remains robust during iterations.

\subsection{Random Network Experiment}
\begin{figure}[H]
\centerline{\includegraphics[width=0.4\textwidth]{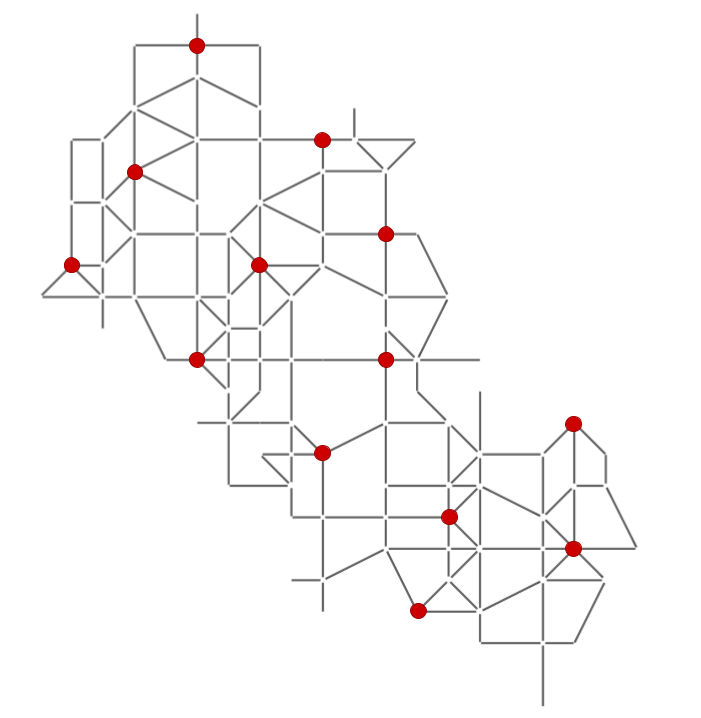}}
\caption{The structure of a random network with OD locations.}
\label{fig:random-grid}
\end{figure}

A Random network was generated in SUMO using netgenerate. This network consists of 348 links and 100 junctions. Each link has a minimum length of 90 and a maximum of 230 meters, with one lane. The structure of the random network is illustrated in Figure \ref{fig:random-grid} with the locations of origins and destinations highlighted in red. A random traffic demand of $8,012$ vehicles was generated for a one-hour simulation period, divided into four 15-minute time intervals.

\begin{figure*}[h] 
     \centering
     \begin{subfigure}[b]{0.47\textwidth}
         \centering
         \includegraphics[width=\textwidth]{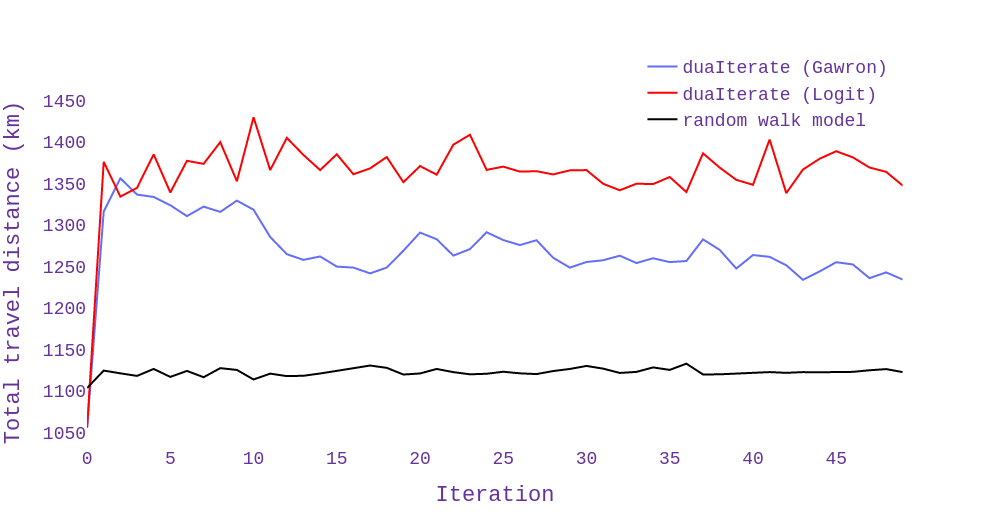}
         \caption{Total travel distance.}
         \label{fig:random-dis}
     \end{subfigure}
     \hfill
     \begin{subfigure}[b]{0.47\textwidth}
         \centering
         \includegraphics[width=\textwidth]{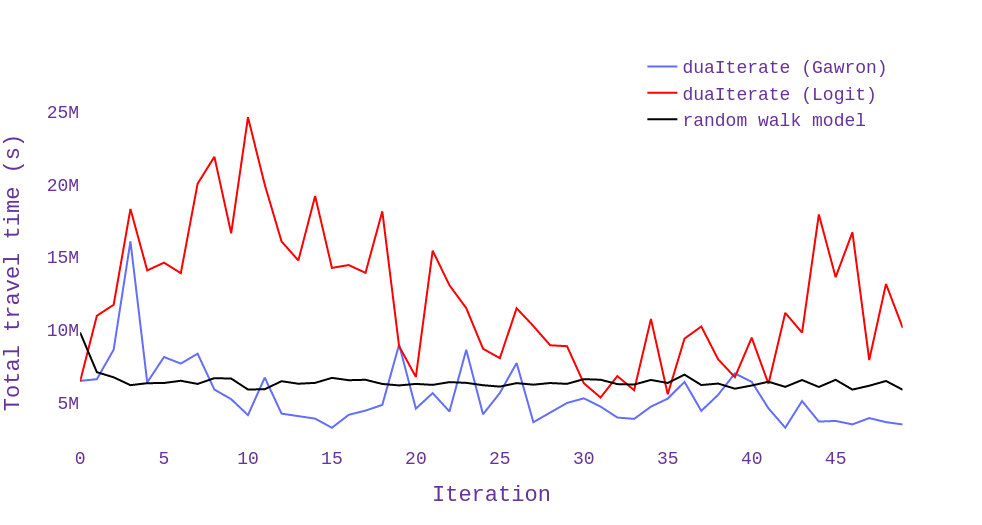}
         \caption{Total travel time.}
         \label{fig:random-time}
     \end{subfigure}
    \caption{Comparison of total travel time and distance for traffic assignment of the proposed method and \textit{duaIterate.py} in the random network.} \label{fig:random-results}
\end{figure*}

Figure \ref{fig:random-results} presents the performance metrics of the proposed route choice model compared to the logit and Gawron models in \textit{duaIterate.py}, for the random network. The results of the random walk model show a considerable improvement in the total travel distance compared to both the logit and Gawron models. The comparison of the total travel time of simulated trips in different iterations demonstrates that the random walk method converges to the user equilibrium condition much faster and remains robust during iterations. However, the total running time of the Gawron model decreases in later iterations.

\begin{figure*}[h!]
     \centering
     \begin{subfigure}[b]{0.30\textwidth}
         \centering
         \includegraphics[width=\textwidth]{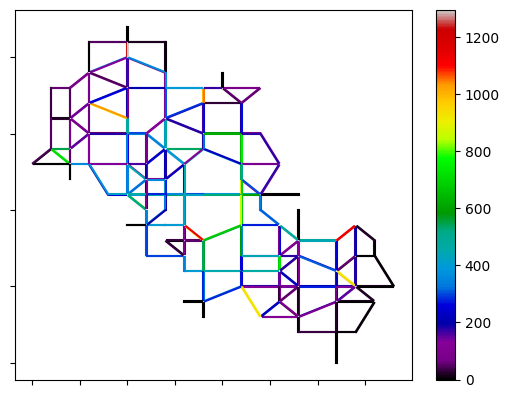}
         \caption{Logit.}
         \label{fig:Heat-logit}
     \end{subfigure}
     \hfill
    \begin{subfigure}[b]{0.30\textwidth}
         \centering
         \includegraphics[width=\textwidth]{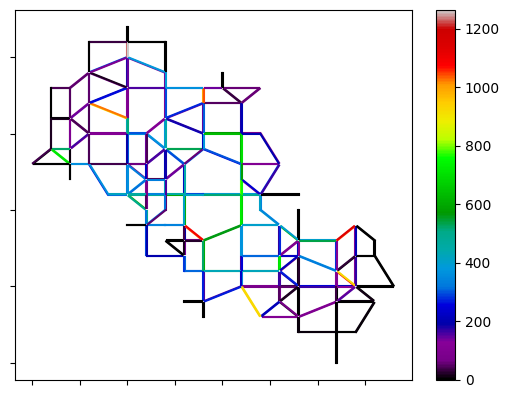}
         \caption{Gawron.}
         \label{fig:Heat-Gawron}
     \end{subfigure}
     \hfill
     \begin{subfigure}[b]{0.30\textwidth}
         \centering
         \includegraphics[width=\textwidth]{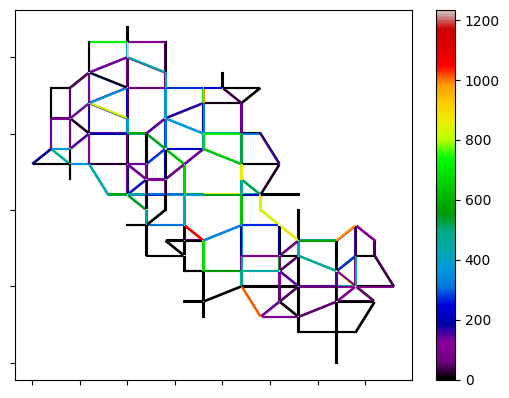}
         \caption{Random walk.}
         \label{fig:Heat-New}
     \end{subfigure}
     \\
    \begin{subfigure}[b]{0.30\textwidth}
         \centering
         \includegraphics[width=\textwidth]{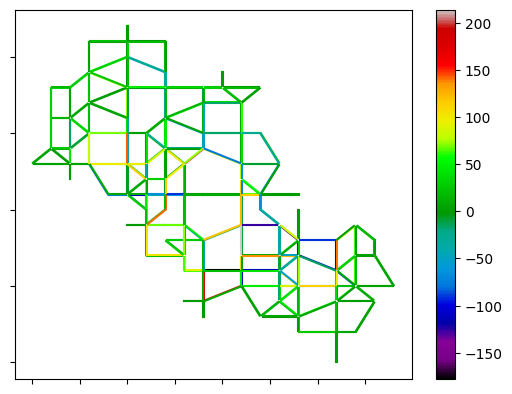}
         \caption{Link count difference \\between logit and Gawron \\models.}
         \label{fig:Logit-Gawron}
     \end{subfigure}
     \hfill
    \begin{subfigure}[b]{0.30\textwidth}
         \centering
         \includegraphics[width=\textwidth]{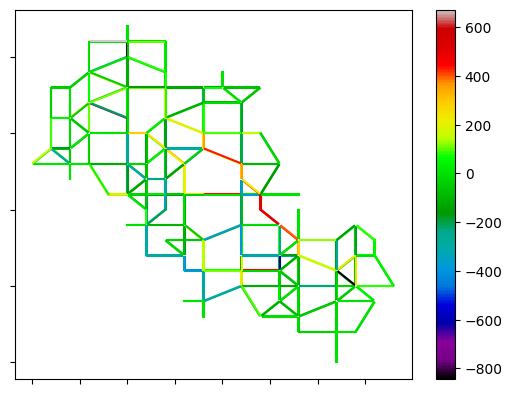}
         \caption{Link count difference \\between random walk and Logit models.}
         \label{fig:New-Logit}
     \end{subfigure}
     \hfill
     \begin{subfigure}[b]{0.30\textwidth}
         \centering
         \includegraphics[width=\textwidth]{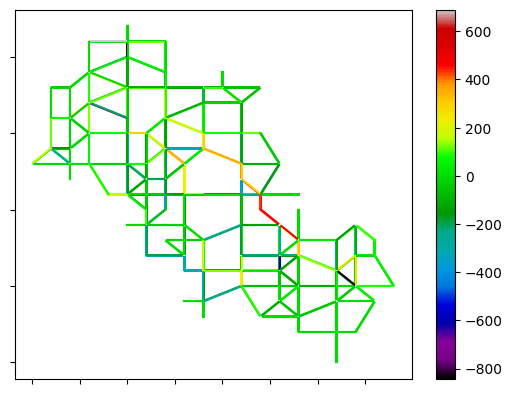}
         \caption{Link count difference \\between random walk and Gawron models.}
         \label{fig:New-Gawron}
     \end{subfigure} 
    \caption{Top row shows the heat maps of traffic distribution using different models. The bottom row shows the difference in link traffic count between each pair of models.} \label{fig:heatmaps}
\end{figure*}

Figure \ref{fig:heatmaps} shows the traffic volume in the random network using different methods. Comparing Figuses \ref{fig:Heat-logit}, \ref{fig:Heat-Gawron}, and \ref{fig:Heat-New}, it is evident that in the random walk method, the number of medium-volume links is slightly higher compared to the logit and Gawron methods. This indicates that the network's traffic volume is more evenly distributed with the random walk method. Figures \ref{fig:Logit-Gawron}, \ref{fig:New-Logit}, and \ref{fig:New-Gawron} highlight the differences in link counts between each pair of models over the entire simulation period. It can be observed that the logit and Gawron models exhibit similar performance, while the random walk model utilizes the central links more frequently. However, the link traffic counts do not significantly differ from the logit and Gawron models.

\section{Conclusion and Future Works}
DTA is a crucial step in providing a realistic representation of traffic conditions in traffic simulation tools. This study introduces a novel dynamic stochastic traffic assignment method that outperforms existing approaches, such as \textit{duaIterate.py}. By calculating route choice probabilities at the intersection level, the model allows for the enumeration of all alternative routes and enables a fully dynamic route assignment process.

The method has been successfully integrated into the SUMO traffic simulation framework and thoroughly evaluated on synthetic networks. All source codes and data are made publicly available, facilitating future research and development within the community.

The results demonstrate its superiority over \textit{duaIterate.py}, showcasing faster convergence, enhanced robustness, realistic trip distribution, and reduced total travel distance. Additionally, the method's ability to execute in parallel significantly reduces running times, making it well-suited for calibration and real-time applications.
The incorporation of randomness and robustness in stochastic approaches is a major concern for DTA solutions. As evident from the results, the proposed method exhibits considerably better robustness compared to \textit{duaIterate.py}.

Furthermore, the average running time of the proposed random walk model in the random network of Figure \ref{fig:random-grid} is approximately 30 seconds per iteration, while \textit{duaIterate.py} models take 27 seconds. Although the primary contributor to the running time is related to running simulations, the convergence rate and robustness of the proposed model make it a promising choice for large real-world networks. 

The proposed dynamic stochastic traffic assignment method provides a valuable contribution to the field of traffic simulation and modeling. Its effectiveness in capturing realistic traffic behavior and optimizing route assignments enhances the understanding and management of transportation systems. Future research can focus on applying the method to real-world networks and exploring further improvements to enhance its scalability and applicability in complex urban environments.


\section{Acknowledgements}
This work is supported by the European Social Fund via IT Academy programme and the Estonian Centre of Excellence in IT (EXCITE). 

We would like to extend our heartfelt thanks to Jakob Erdmann for his fast and insightful response to our SUMO-related technical questions. His assistance has been invaluable in furthering our research objectives.

Language Model GhatGPT-3.5 is used for proofreading this document.  


 \section{AUTHOR CONTRIBUTIONS}

 The authors confirm contribution to the paper as follows: study conception and design: K. Khoshkhah, M. Pourmoradnasseri; data and implementation: K. Khoshkhah; analysis and interpretation of results: K. Khoshkhah, M. Pourmoradnasseri; draft manuscript preparation: K. Khoshkhah, M. Pourmoradnasseri, S. Ben Yahia, A.Hadachi. All authors reviewed the results and approved the final version of the manuscript.

\bibliographystyle{trb}
\bibliography{trb_template}

\end{document}